\documentclass[twocolumn,showpacs,pre,final]{revtex4}
\usepackage{amscd}
\usepackage{amssymb}
\usepackage{amsfonts}
\usepackage{amsmath}
\usepackage{graphicx}
\usepackage{dcolumn}
\usepackage[dvipdfm]{hyperref}

\hypersetup{colorlinks=true, linkcolor=blue, anchorcolor=blue,
citecolor=blue, filecolor=blue, pagecolor=blue, urlcolor=blue}

\begin{document}

\preprint{}
\title{Reciprocal relationships in collective flights of homing pigeons}
\author{Xiao-Ke Xu$^{1,2}$}
\email{xiaokeeie@gmail.com}
\author{Graciano Dieck Kattas$^1$}
\author{Michael Small$^{3,1}$}
\affiliation{$^1$Department of Electronic and Information
Engineering, Hong Kong Polytechnic University, Hung Hom, Kowloon,
Hong Kong \\$^2$School of Communication and Electronic Engineering,
Qingdao Technological University, Qingdao 266520, China \\$^3$School
of Mathematics and Statistics, University of South Australia, Mawson
Lakes, SA 5095, Australia}

\date{\today}

\begin{abstract}
Collective motion of bird flocks can be explained via the hypothesis
of many wrongs, and/or, a structured leadership mechanism. In
pigeons, previous studies have shown that there is a well-defined
hierarchical structure and certain specific individuals occupy more
dominant positions --- suggesting that leadership by the few
individuals drives the behavior of the collective. Conversely, by
analyzing the same data-sets, we uncover a more egalitarian
mechanism. We show that both reciprocal relationships and a
stratified hierarchical leadership are important and necessary in
the collective movements of pigeon flocks. Rather than birds
adopting either exclusive averaging or leadership strategies, our
experimental results show that it is an integrated combination of
both compromise and leadership which drives the group's movement
decisions.
\end{abstract}

\pacs{89.75.Fb, 89.75.Kd, 89.75.Da}


\maketitle


\section{Introduction}
The fast coherent movement of flocking birds is a fascinating
phenomenon exhibiting apparent intelligence and coordination
\cite{Potts1984}. New monitoring technologies have meant that this
collective behavior has recently attracted renewed interest from
scientific and engineering communities, allowing more conclusive
analysis to be performed \cite{Organized_flight, Collective_motion}.
It is a valuable topic whether all members of a flock contribute
equally to the collective decision making and follow equivalent
rules, or certain individuals have a greater influence on the
decisions of the group \cite{Collective_motion,
Group_decision-making, Haitao_paper,Hierarchical_group,
Leadership_pigeon, Modelling_Group_Navigation, Freeman_2010}.

The proposition that all contribute equally is sometimes referred to
as the ``many wrongs" principle, and purports that individuals
average their preferred directions depending on interaction with
their neighbors, leading to a compromise in route choice
\cite{Many-wrongs_principle1, Many-wrongs_principle2}. Conversely,
the leadership hypothesis posits that one or a small number of
leaders are able to exert a disproportionate influence on the
group's movement decisions \cite{Leadership_fish,
Effective_leadership}. Both theoretical and experimental arguments
predict that the compromise of all members will make more accurate
decisions than the leading of one or a small number individuals,
unless leaders have very different and superior information
\cite{Group_decision-making}. For example, homing performance of the
pre-trained pigeons flying as a flock is significantly better than
that of these birds released individually \cite{Group_efficient}.
Yet, recent research has shown that time-varying hierarchical
decision making mechanisms do exist during pigeon flights
\cite{Collective_motion, Hierarchical_group}
--- giving strength to the leadership hypothesis.

Although new studies show that certain individuals in pigeon flocks
are able to exert relatively more influence on the movement
decisions of the whole group, only the directed relationship
(pointing from the leader to the follower) has been studied
\cite{Leadership_pigeon, Hierarchical_group}. In addition to such
directed links (representing a leader-follower relationship)
selective coordinated behavior may also exist in collective motion,
hence mutual relationships must also be examined. For example, in
the case of a perturbation caused by the terrain or a predator, the
better strategy for birds would probably be to share the information
of all group members to move rapidly to safety regardless of their
individual positions within a leadership hierarchy. Such mutual
links represent a reciprocal relationship between a pair of pigeons,
which appears nonrandomly in real-life directed networks
\cite{Definition_reciprocity, Pattern_reciprocity} and plays a
significant role in the evolution of many biological systems
\cite{Evolution_reciprocity, Indirect_reciprocity}. We emphasize
that the coexistence of compromise and leadership is not a
contradiction, but a meaningful supplement to the hierarchical
structure of pigeon flocks. The main difference between the work in
\cite{Hierarchical_group} and ours is as follows: they are
concentrating on the leadership aspect; while we are focusing on
local interactions (including both directed and reciprocal
relationships).

In this paper, we re-analyze experimental high-precision datasets of
pigeon flocks to arrive at a more nuanced conclusion about the
interactions and decisions in the collective dynamics of birds.
Using quantitative methods from statistical physics
\cite{Hierarchical_group}, we find that both outcomes (directed and
mutual links) coexist in the same flock flights. The mutual links
represent a reciprocal relationship between individuals, which is a
useful supplement to the well-defined hierarchical structure.
Integrating both directed and reciprocal links we uncover the
complete topology of the network induced by the collective motion of
a pigeon flock. Most significantly, our results imply that there is
an integrated mechanism of decision-making in pigeon flocks: neither
a leadership nor a compromise mechanism is clearly dominant, rather
both mechanisms coexist.

\section{Results}
\subsection{Reciprocal relationships by calculating pairwise correlations}
In the past tens of years, it is a very difficult mission to explore
the influence of individual members on a fast collective motion at
all times. Recently, the advance of GPS devices allow us to use
sophisticated evaluation techniques to mine real flocking data
\cite{Collective_motion}. Employing high-precision GPS in tracking
pairs of pigeons, Biro \emph{et al.} find that two birds compromise
if they have less diversity on directional preferences, while either
the pair split or one of them becomes the leader for a severe
conflict \cite{Leadership_pigeon}. Using lightweight GPS devices and
analyzing data concerning leading roles in pairwise correlations,
Nagy \emph{et al.} show a well-defined hierarchy among pigeons
belonging to the same flock \cite{Hierarchical_group}. In this
study, we use the same datasets: $11$ free flights and $4$ homing
flights. More detailed information on the datasets can be obtained
from the website:
\href{http://hal.elte.hu/pigeonflocks/}{http://hal.elte.hu/pigeonflocks/}.

To investigate the influence that a given bird's behavior has on the
other flock members, the temporal relationship between the flight
directions has been evaluated \cite{Hierarchical_group,
Collective_motion}. The directional correlation for a pair of
pigeons is $C_{ij}(\tau)=\left <\overrightarrow{V_i}(t)\cdot
\overrightarrow{V_j}(t+\tau)\right >$, where $\left <\cdots \right
>$ denotes time average and $\overrightarrow{V_i}(t)$ is the
normalized velocity of bird $i$. When $C_{ij}(\tau)$ obtains its
maximum value at the time delay $\tau_{ij}$, $\tau_{ij}$ is called
the optimal directional delay time. Negative $\tau_{ij}$ values mean
that the $i$th bird falls behind the $j$th bird, which can thus be
interpreted as a case of $j$ leading. For each pair, we extract the
positive value $\tau_{ij}=-\tau_{ji}$ as a directed edge pointing
from the leader to the follower.

\begin{figure}[htbp]
\centering
\includegraphics[width=0.5\textwidth]{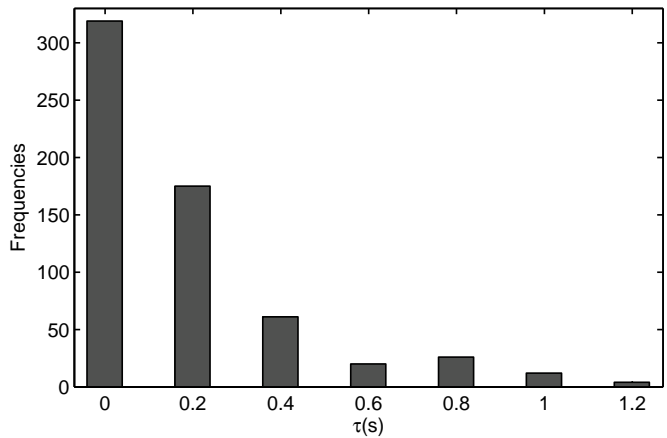}
\caption{Histogram for the different values of the directional
correlation delay time $\tau_{ij}$ between each pair of pigeons in
all the flights. When the correlation coefficient reaches the
maximum value $C_{ij}(\tau)$ at the time delay $\tau_{ij}$,
$\tau_{ij}$ is called the directional correlation delay time and
$C_{ij}(\tau)$ is called the maximum correlation coefficient. Here
we only consider the conditions of $\tau_{ij}\geq 0$. }
\label{fig1_histogram}
\end{figure}

If individuals fly together in a flock, they will show a very
similar velocity and a high correlation \cite{Scalefree_PNAS}. In a
previous study, the authors study the leader-follower relationship
among pairs of pigeons whose directional correlation time delay is
non-zero, and such links are directed \cite{Hierarchical_group}. The
directed link indicates that a following bird tends to consistently
copy the directional behavior of particular leading individuals.
However, the directional correlation time delay may be near zero,
which means a pair of pigeons have a coordinated interaction with
one another and there is a mutual (reciprocal) link between them.
The frequency distribution of the directional correlation time delay
for pairwise pigeons in all collective motions is shown in
Fig.~\ref{fig1_histogram}. Here we select $0.2$ s as the time
interval of time delay, for the sampling time interval of the
original dataset and the resolution of time delay in the previous
study \cite{Hierarchical_group} both are $0.2$ s. The frequency of
$\tau=0$ is the most frequent, meaning that there are typically many
mutual relationships between the pairwise birds in pigeon flocks.
Moreover, the result also implies that the perfect hierarchical
structure induced by directed links in \cite{Hierarchical_group} may
not be adequate to completely explain the collective behavior of
pigeon flocks.

\begin{figure}[htbp]
\centering
\includegraphics[width=0.5\textwidth]{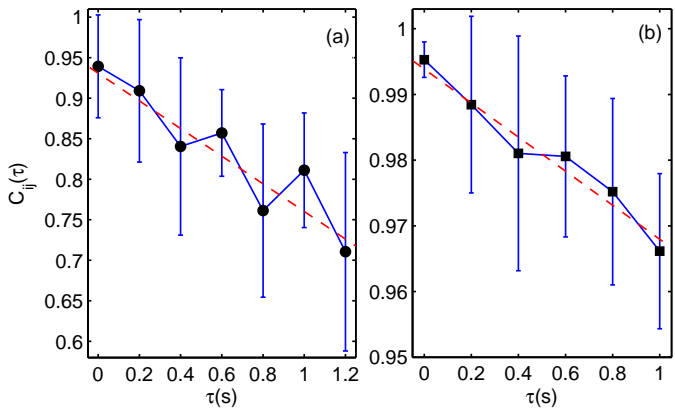}
\caption{The values of $C_{ij}(\tau)$ decrease with $\tau_{ij}(s)$
for all the (a) free flights and (b) homing flights. The solid
circles (squares) show the mean of $C_{ij}(\tau)$ while the bars
show the standard deviation of $C_{ij}(\tau)$.} \label{fig2_Cij}
\end{figure}

Here we hope to explain why a correlation that decays from $\tau=0$
represents a ``mutual link'' instead of no interaction in our study.
Actually, in some cases of collective motions, a decaying temporal
correlation that is maximal at zero lag is an evidence that there is
no interaction. For example, Katz and collaborators showed that the
orientation correlation whose peak at zero time delay is
significantly lower than the orientation correlation whose peak
after zero [Fig. S8A in ~\cite{Katz_2011}]. On the contrary, in our
study $C_{ij}(\tau)$ at $\tau=0$ tends to be higher than the values
at the large time delay [Fig.~\ref{fig2_Cij}], which means the
strength of mutual links is stronger than that of directed links.

Furthermore, we use multiple effective methods ($C_{ij}(\tau)$,
motif, synchronization and multiscale analysis) to prove the
existence of reciprocal interaction as follows. The values of the
maximum correlation coefficient $C_{ij}(\tau)$ at the directional
correlation time delay $\tau_{ij}$ have been shown in
Fig.~\ref{fig2_Cij}. The birds' movements under two conditions, free
flights and homing flights, are recorded~\cite{Hierarchical_group}.
The pigeon flock makes a circle-like route in free flights; while
the group makes less direction turning during homing flights [Fig.
S6 in \cite{Hierarchical_group}]. Because of the more centralized
distribution of turning directions in homing flights, it is easier
to obtain a larger $C_{ij}(\tau)$, which results in the values of
$C_{ij}(\tau)$ at the same $\tau_{ij}$ in homing flights
[Fig.~\ref{fig2_Cij}(b)] being larger than those in free flights
[Fig.~\ref{fig2_Cij}(a)]. In both cases, we can find that
$C_{ij}(\tau)$ at the small $\tau_{ij}$ tends to be higher than the
values at the large time delay. Moreover, when the time delay is
zero, the value of $C_{ij}(\tau)$ is the highest, which implies that
these mutual links are the most important relationship in the flock
and more attention should be paid to such types of links.

Although interaction and correlation are different and their
relationship has been extensively discussed \cite{Scalefree_PNAS},
using the method of calculating pairwise correlation
\cite{Hierarchical_group}, it is still difficult to determine
whether the relationship between a pair of pigeons is a direct
interaction or an indirect correlation. Here we develop a simple
method to detect whether three links in a small subgraph (motif) are
independent. A motif, defined as a small connected subgraph that
recurs in a graph, is the basic functional unit of complex networks
\cite{Motif, Timeseries_pnas}. A small motif with three nodes in the
network induced from each flight are shown in
Table~\ref{table_motif}.

\begin{table}[htbp]
\centering \caption{Comparison of the values of the maximum
correlation coefficient for the three links in the subgraph (motif)
structure. $C_{1}$, $C_{2}$ and $C_{3}$ represent the maximum
correlation coefficients for the three links, respectively. $n$ is
the whole number of the subgraph in all the flights.}
\begin{ruledtabular}
\begin{tabular}{c c c c c c}
\raisebox{-2.0ex}[0pt]{Motif} & \raisebox{-2.0ex}[0pt]{$n$} &
\raisebox{-2.0ex}[0pt]{$C_{1}>C_{2}$} &
\raisebox{-2.0ex}[0pt]{$C_{1}>C_{3}$} &
\raisebox{-2.0ex}[0pt]{$C_{2}>C_{3}$}
 & $C_{1}>C_{3}$ \\
 & & & & & $C_{2}>C_{3}$ \\\hline
\includegraphics[width=0.9in]{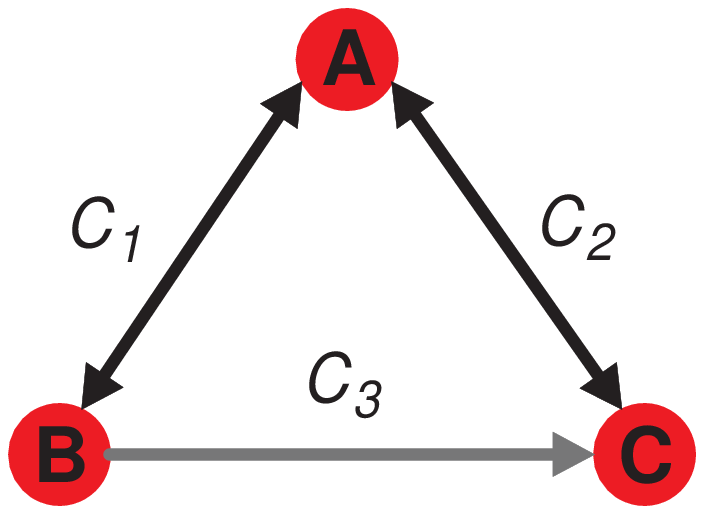} & \raisebox{4.5ex}[0pt]{$173$} & \raisebox{4.5ex}[0pt]{$54\%$} & \raisebox{4.5ex}[0pt]{$69\%$} & \raisebox{4.5ex}[0pt]{$64\%$} & \raisebox{4.5ex}[0pt]{$50\%$} \\
\end{tabular}
\end{ruledtabular}
\label{table_motif}
\end{table}

If the relationships among the three individuals are independent,
the probabilities of $C_{1}>C_{2}$, $C_{1}>C_{3}$, and $C_{2}>C_{3}$
should be about $50\%$ respectively, while $C_{1}>C_{3}$ along with
$C_{2}>C_{3}$ simultaneously should have a probability of $33\%$.
However, we observe that all these relationships (except
$C_{1}>C_{2}$) occur significantly more frequent than we would
expect [Table~\ref{table_motif}], which means that $B$ and $C$ tend
to have a lower correlation than that between $A$ and $B$, and
between $A$ and $C$. Hence, the correlations among the three
individuals are not consistent with three independent events and the
directed link between $B$ and $C$ ($C_3$) can be attributed to the
other two high mutual correlations ($C_{1}$ and $C_{2}$). Again, our
result implies that we can not neglect mutual links, or believe that
the nodes with such type of relationship are no interaction.

\subsection{Multi-scale analysis of reciprocal links}
\begin{figure}[htbp]
\centering
\includegraphics[width=0.5\textwidth]{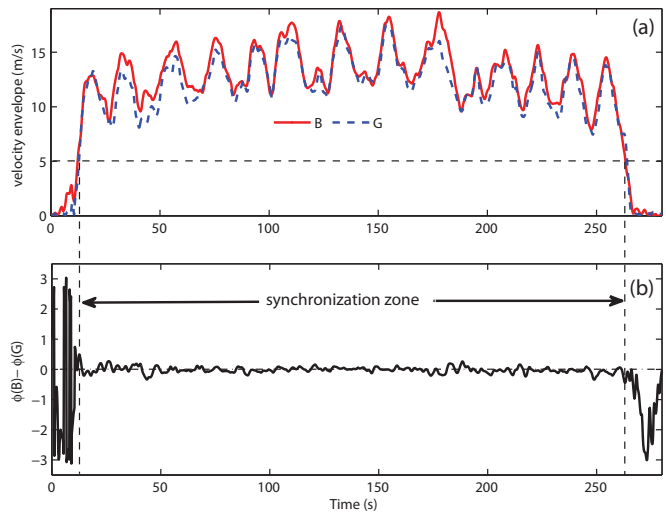}
\caption{A pair of birds (B and G) in the eleventh free flight show
a very strong synchronization ($\tau_{ij}=0$) and correlation (
$C_{ij}(\tau)=0.91$) at the time scale from $10$-$270$s. (a)
Velocity envelopes (amplitudes) of B and G, and (b) phase
differences between the pair of birds.} \label{fig3_synchronization}
\end{figure}

Although the pioneering researchers have obtained the detailed
spatiotemporal data on the positions of individuals during group
movements, they do not use the data to study the dynamical variation
of the relationship in collective motions \cite{Hierarchical_group}.
To make clear the spatiotemporal relationship for a pair of
mutual-link birds, we show the synchronization of B and G in the
eleventh free flight [Fig.~\ref{fig3_synchronization}]. The high
synchronization of two pigeons again provides evidence that those
pairs with mutual links ($\tau_{ij}=0$) are strongly interacting
rather than completely independent. The above result implies that
the consensus in a real-life bird flock can also be achieved by the
neighbor compromise mechanism, if we do not consider they might
``copy'' the flight direction of their common leader.

\begin{figure}[htbp]
\centering
\includegraphics[width=0.5\textwidth]{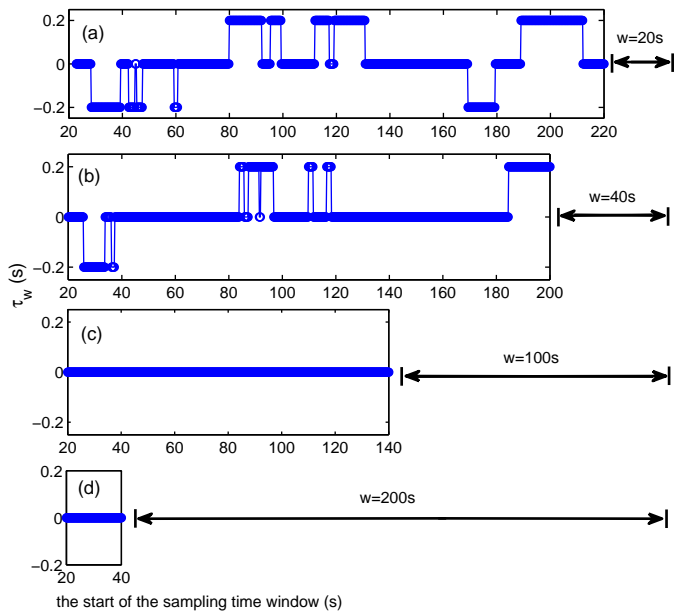}
\caption{Multi-scale pairwise correlation analysis for the different
time-window series of B and G in the eleventh free flight. When the
directional correlation of each time window for a pair of pigeons
obtains its maximum value at the time delay $\tau_w$, $\tau_w$ is
called the directional correlation delay time of each time window.
For the whole trajectories of B and G, the maximum correlation
coefficient $C=0.91$ and the optimal time delay $\tau=0$. The
sampling time interval and the resolution of time delay both are
$0.2$ s. Because the pigeons make a circle-like flight in free
flights and their one-dimensional flight trajectory is like a
pseudo-periodic time series \cite{Michael_PRL, Zhangjie_PRL}, we
select the approximate cycle ($w=20$ s) as the minimum time window.
Pairwise correlation analysis for (a) the time window is $20$ s. We
only calculate the optimal time delay for the time scale from
$20$-$240$ s (the strong synchronization segment). The x axis is the
start time of the sampling window. When the start time of the
sampling window is $20$ s, $\tau_w(t=20.0)$ is obtained by
calculating the directional correlation during the period from
$20.0$ s to $40.0$ s. Next value $\tau_w(t=20.2)$ is obtained from
the period between $20.2$-$40.2$ s, and so on up to the interval
$220.0$-$240.0$ s. (b) the time window is $40$ s, (c) the time
window is $100$ s, and (d) the time window is $200$ s. }
\label{fig4_multiscale}
\end{figure}

The collective movement of birds, such as during the abrupt
splitting of a flock, can instantaneously change. Therefore, we
first divide a long time series into many short overlapping
segments, and then study the pairwise correlation on each segment.
Changing the scale of the time window, we can get the dynamical
$\tau_{ij}$ for multiple timescales. The variation of the
relationship of individuals B and G on multiple timescales in a
flight has been shown in Fig. \ref{fig4_multiscale}. Because the
pigeons make a circle-like flight in the free flights and their
one-dimensional flight trajectory is like a pseudo-periodic time
series \cite{Michael_PRL, Zhangjie_PRL}, we select the approximate
cycle ($20$ s) as the minimum time window. The relationship between
B and G varies fast with time at the small timescale [Fig.
\ref{fig4_multiscale}(a)]. Our result shows that the collective
behavior of the birds varies with a short timescale, and the
relationship between a pair of pigeons is time dependent. However,
With the timescale increasing, the frequency with which the
relationship varies reduces, for the result is an average of a
longer time [Figs. \ref{fig4_multiscale}(b)-(d)]. Obviously, the
longer the time averaging is, the stabler the result of the
correlation function becomes.

Recent work has illustrated the group decision rule for homing
pigeons: compromise for small conflict and leadership for large
\cite{Leadership_pigeon}. In this study, it is difficult to measure
the conflict level among pigeons in each flight. Nevertheless, we
find the compromise and leadership both emerge in the \emph{spatial}
domain from each dataset by calculating pairwise correlation
[Fig.~\ref{fig5_network} and Table~\ref{table_all}]. However, we do
not find the existence of a stable leader-follower relationship
between a pair of pigeons throughout all the flights, and this
result is different to the findings in \cite{Leadership_pigeon}. For
example, two pigeons of A and B do not show the same relationships
(different sign of $\tau(s)$) in all the flights [Table.
\ref{table_AB}]. A leads B for three flights, A follows B for five
flights, and there is no leadership between them for three flights.
That means, considering a very long time scale (e.g., all the
flights in this study), there is no single individual always leading
a pair of individuals. Actually, A leading B in this flight and then
B leading A in next time maybe can be regarded as a general concept
of reciprocal behaviors (``You scratch my back, and I'll scratch
yours'' \cite{Indirect_reciprocity}). The relationship switching
among group members supports more flexibility for individuals to
response to external predators \cite{Pomeroy_1992}.

\begin{table*}[htbp]
\centering \caption{The correlation delay time $\tau(s)$ and the
corresponding correlation coefficient $C(\tau)$ between A and B in
all the flights. FF means free flight and HF means homing flight.
Here ``$-$'' represents that the individual does not attend the
flight, so we have no data for analyzing its behavior.}

\smallskip

\begin{ruledtabular}
\begin{tabular}{c c c c c c c c c c c c c c c c}
Flight &FF1 & FF2 & FF3 & FF4 & FF5 & FF6 & FF7 & FF8 & FF9 & FF10 & FF11 & HF1 & HF2 & HF3 & HF4 \\
\hline
$\tau(s)$ & $-$ & $0$ & $-0.4$ & $0$ & $1.0$ & $-$ & $0$ & $-$ & $0.2$ & $-0.2$ & $-0.6$ & $-$ & $-0.2$ & $0.2$ & $-0.2$ \\
$C(\tau)$ & $-$ & $0.98$ & $0.75$ & $0.88$ & $0.86$ & $-$ & $0.95$ & $-$ & $0.73$ & $0.84$ & $0.84$ & $-$ & $0.99$ & $1.00$ & $0.98$ \\
\end{tabular}
\end{ruledtabular}
\label{table_AB}
\end{table*}

\subsection{Coexistence of compromise and leadership}
\begin{figure}[htbp]
\centering
\includegraphics[width=0.5\textwidth]{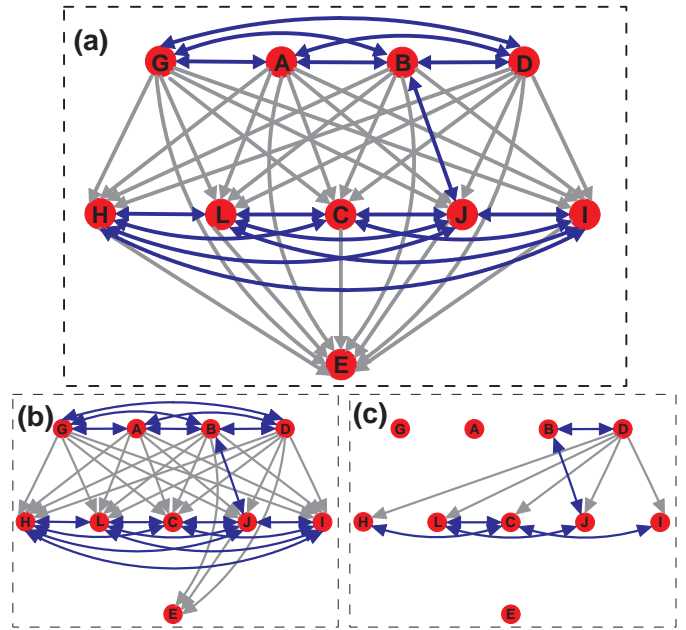}
\caption{The network topology of the second free flight. If the
directional correlation time delay $\tau_{ij}>0$ and the maximum
correlation coefficient $C_{ij}(\tau)>C_{min}$, we establish a
directed link for the network. The gray arcs represent directed
links from leaders to followers. If $\tau_{ij}=0$ and
$C_{ij}(\tau)>C_{min}$, we build a mutual link. The blue lines are
mutual links, which represent a coordinated correlation between a
pair of equal pigeons. (a) We construct the network by including
only those edges whose maximum correlation values $C_{ij}(\tau)$ are
above a given variable minimum, $C_{min}=0.5$. E is the follower of
all the other individuals. (b) Selecting other thresholds (such as
$C_{min}=0.9$) to maintain the network connectivity throughout the
pigeon flock, also gives similar results. (c) The disconnected
network is built by choosing $C_{min}=0.99$. Maintaining the network
connectivity throughout the flight of the pigeon group is a key
issue \cite{Su_2010}. Therefore, it is not suitable to select a very
high threshold to let the network lose the connectivity for the
pigeon group.} \label{fig5_network}
\end{figure}

We have shown that mutual links are ubiquitous and important in
collective flights, so it is very valuable to investigate whether
the conclusion that certain leaders are able to exert more influence
on the group's movement decisions still holds
\cite{Hierarchical_group, Collective_motion}. We construct a flight
network by including only those links whose maximum correlation
values $C_{ij}(\tau)$ are above a given variable minimum,
$C_{min}=0.5$ [Fig.~\ref{fig5_network}(a)]. Here E is the follower
of all the other individuals. Selecting different thresholds in a
suitable range ($C_{min}\in [0.5-0.95]$) to maintain the network
connectivity gives similar results [Fig.~\ref{fig5_network}(b)]. In
such a network the nodes represent individual birds, while the links
(arcs) denote inferred relations between their movements. If
$C_{ij}(\tau)>C_{min}$ and $\tau_{ij}>0$, we build a directed link
pointing from the leader $i$ to the follower $j$. While if
$C_{ij}(\tau)>C_{min}$ and $\tau_{ij}=0$, we generate a mutual link
for the pair of pigeons. Hence we have two types of links in the
network induced from the collective movement of a pigeon flock:
directed and mutual links.

If we consider only directed links, a well-defined hierarchical
structure is evident. On the other hand, considering only mutual
links, we find that most individuals share an equal structure and
the individuals with reciprocal links form many loops. Finally, when
we take all the links into account, the topological structure is
very complex because not only many directed links compose a
hierarchical structure but also many reciprocal links form a large
range of equal relationships. Instead of showing birds exclusively
adopting either an averaging or a leadership strategy, our
experimental analysis demonstrates that there is an integrated
mechanism between compromise and leadership.

Our data analysis shows that mutual (reciprocal) links, representing
a coordinated correlation between a pair of individuals, are dense
and important in pigeon flocks. A traditional way of quantifying the
reciprocity is to compute the ratio of the number of links pointing
in both directions $L^\leftrightarrow$ to the total number of links
$L$\cite{Definition_reciprocity}: $l=\frac{L^\leftrightarrow}{L}$.
In general, the reciprocity of real-life directed networks ranges
between the two extremes of a purely directed one ($l=0$, such as
citation networks, where recent papers can cite less recent ones
while the opposite cannot occur) and of a purely bidirectional one
($l=1$, such as the Internet, where information always travels both
ways along computer cables) \cite{Pattern_reciprocity}. The value of
$l$ for a real network lies between the above two extremes. As is
shown in Table~\ref{table_all}, there are many mutual links and the
reciprocal coefficient $l$ is very high in each network of the
homing and free flights. Moreover, the high frequency of loops (the
size of $3$) $m$ shows that not only reciprocal relationships exist
between two individuals but also such type of transitivity \cite{SW,
Newman_review} can be extended to three or a larger number of
individuals.

The number of individuals with reciprocal links and the entire
number of individuals in a flight have been listed in
Table~\ref{table_all}, and we find that there are very few pigeons
having no reciprocal links. Furthermore, we list the ranks of
individuals without reciprocal links, and we find such pigeons tend
to be a pure leader or follower when observing their ranks in the
group, like $E$ in Fig.~\ref{fig5_network}. Therefore, a reasonable
conclusion is that the individual whose position is the head or tail
of the flock has a stronger tendency to have a directed relationship
with others. In contrast, the individuals in the middle of the group
tend to have a reciprocal correlation with their neighbors. We
emphasize that the hierarchical structure of pigeon flocks in the
previous study \cite{Hierarchical_group} does not imply that there
are no mutual correlations: they simply concentrate on the
leadership aspect. Our findings show that both the many wrongs and
leadership mechanism coexist in collective motion. Within and
between distinct strata, the hierarchical structure likely dominates
long term decisions such as navigation objectives of the whole
flock, while the mutual interactions characterize local behavioral
rules that are essential to maintain flock cohesion and alignment.

\begin{table*}[htbp]
\centering \caption{Summary of the statistics in each free and
homing flight performed by subjects. $L^\leftrightarrow$ is the
number of mutual links, $L$ is the total number of links, $l$ is the
ratio of the number of links pointing in both directions
$L^\leftrightarrow$ to the total number of links $L$:
$l=\frac{L^\leftrightarrow}{L}$ \cite{Definition_reciprocity}, $m$
is the number of loop of size $3$, $N^\leftrightarrow$ is the number
of pigeons with the reciprocal relationship, $N$ is the total number
of pigeons in the flight, $D$ is the serial number of the
individuals without reciprocal links to others, and $R$ is the ranks
of $D$ in each flight. ``$-$'' represents that the individual does
not attend the flight, so we have no data for analyzing its
behavior.}

\smallskip

\begin{ruledtabular}
\begin{tabular}{c c c c c c c c c c c c c c c c}
Flight &FF1 & FF2 & FF3 & FF4 & FF5 & FF6 & FF7 & FF8 & FF9 & FF10 & FF11 & HF1 & HF2 & HF3 & HF4 \\
\hline
$L^\leftrightarrow$ & $16$ & $34$ & $20$ & $14$ & $14$ & $24$ & $44$ & $42$ & $14$ & $36$ & $26$ & $20$ & $12$ & $22$ & $14$ \\
$L$ & $34$ & $62$ & $31$ & $32$ & $52$ & $34$ & $50$ & $49$ & $35$ & $54$ & $58$ & $55$ & $42$ & $39$ & $43$ \\
$l$ & $0.47$ & $0.55$ & $0.65$ & $0.44$ & $0.27$ & $0.71$ & $0.88$ & $0.86$ & $0.40$ & $0.67$ & $0.45$ & $0.36$ & $0.29$ & $0.56$ & $0.33$ \\
$m$ & $16$ & $35$ & $17$ & $16$ & $8$ & $29$ & $79$ & $70$ & $8$ & $56$ & $41$ & $32$ & $13$ & $25$ & $9$ \\
$N^\leftrightarrow$ & $7$ & $9$ & $7$ & $9$ & $7$ & $8$ & $7$ & $8$ & $7$ & $8$ & $9$ & $7$ & $8$ & $6$ & $7$ \\
$N$ & $8$ & $10$ & $7$ & $9$ & $10$ & $8$ & $7$ & $8$ & $8$ & $9$ & $10$ & $10$ & $9$ & $8$ & $9$ \\
$D$ & H & E & $-$ & $-$ & A,M,G & $-$ & $-$ & $-$ & I & M & A & A,D,I & G & C,L & A,G \\
$R$ & $8$ & $10$ & $-$ & $-$ & $1,2,3$ & $-$ & $-$ & $-$ & $8$ & $9$ & $10$ & $1,3,5$ & $9$ & $7,8$ & $2,9$ \\
\end{tabular}
\end{ruledtabular}
\label{table_all}
\end{table*}

\section{Conclusion and Discussion}
Our study indicates the balance between compromise and leadership
for the organized flight of pigeons. Reciprocal links represent a
mutual correlation between a pair of individuals. Note that our
result is not contradictory to the previous conclusion that there is
a hierarchical structure in pigeon flocks \cite{Hierarchical_group},
instead it is a meaningful supplement. Our results show that the
many wrongs and leadership mechanism can coexist in a collective
motion. Hence, the dichotomy between these two mechanisms is false,
at least for the flocking flight of pigeons.

Our work also has significant meaning for modeling the collective
motion of animals. If a pigeon flock only has a hierarchical
structure, it means that the local interaction mechanism of previous
models \cite{Reynolds1987, Vicsek1995} may not be adequate for
simulating the group flights of homing pigeons (lacking the
leadership), despite the interaction rule being dependent on the
metric distance \cite{Vicsek1995, Experiment_PNAS, Vicsek_2000} or
the topological distance \cite{Topological_distance}. However, our
work suggests that the local interaction relationships (including
directed and mutual links) are sufficient to characterize cohesive
motion of pigeon flocks. Our results are helpful to provide a
comprehensive picture of collective dynamic behavior in animal group
movement and unify both the interaction mechanisms observed in
experimental data and theoretical models of coherent behavior.

We hypothesize that the integrated mechanism between compromise and
leadership also brings more advantage to both the individuals and
the whole system, to cope with external perturbations (e.g.,
predatory threat and food source). Individuals of the same species
come together to form a group because a compact flock has more
advantages to react to environmental perturbations than separate
individuals \cite{Group_efficient, Sumpter_book, Ward_2011}. Any
external perturbation for flocking movement is likely to directly
cause a change of velocity (direction, magnitude, or both) for a
small subset of birds that first detect the perturbation. Such
localized changes can be transmitted to the whole flock to produce a
collective response as if being of one mind \cite{Couzin_2009},
making the whole group both very flexible and responsive.
Conversely, if there were only a hierarchical structure in pigeon
flocks, the followers cannot transmit any external signals to the
leaders, so the whole group is not so sensitive to environmental
perturbations. The sensitivity of the whole group can evolve and
strengthen if the group members have a local interaction mechanism
among them \cite{Sumpter_2008}. In particular, reciprocal links
allow individuals to interact with their neighbors and supply a
useful way for the followers to convey information to the leaders.

It needs to be noted that calculating pairwise correlation and even
the methods in \cite{Leadership_pigeon, Hierarchical_group} can be
regarded as dividing a large-scale complex system (such as a fish
school \cite{Ward_2011, Herbert-Read_2011}) into multiple local
sub-systems (a pair of subjects). It is difficult to accurately
determine how one individual is simultaneously affected by others
(more than one individual), because the whole is greater than the
sum of its parts \cite{More_is_different}. An extension of this
method based on the holism of complex systems \cite{Newman_survey}
for analyzing these trajectory data should be developed in future.

\begin{acknowledgments}
The authors are deeply indebted to Mr. M\'{a}t\'{e} Nagy who has
kindly provided the trajectory data of pigeon flocks. This work was
supported by the PolyU Postdoctoral Fellowships Scheme (G-YX4A) and
the Research Grants Council of Hong Kong (BQ19H). G.D.K. is
currently supported by the Hong Kong PhD Fellowship Scheme from the
Research Grants Council of Hong Kong. X.-K.X. also acknowledges the
National Natural Science Foundation of China (61004104, 61104143).
\end{acknowledgments}


\end{document}